\documentclass[12pt]{article}
\usepackage{graphics,epsfig}
\usepackage[english]{babel}
\newcommand{\cinst}[2]{$^{\mathrm{#1}}$~#2\par}
\newcommand{\crefi}[1]{$^{\mathrm{#1}}$}

\begin{document}

\thispagestyle{empty}
\begingroup


\vglue.5cm \hspace{7cm}{YerPhI Preprint 1623 (2010)
\vglue.5cm
\begin{center}


{\normalsize \bf NUCLEAR EFFECTS IN REACTIONS $\nu p \rightarrow
\mu^- p \pi^+$ and
\\[.3cm]
$\nu p \rightarrow \mu^- \Delta^{++}(1232)$ ON BOUND PROTONS}

\end{center}

\vspace{1.cm}

\begin{center}
{\large SKAT Collaboration}

 N.M.~Agababyan\crefi{1},
 N.~Grigoryan\crefi{2}, H.~Gulkanyan\crefi{2},
 A.A.~Ivanilov\crefi{3},\\ Zh.~Karamyan\crefi{2},
V.A.~Korotkov\crefi{3}

\setlength{\parskip}{0mm}
\small

\vspace{1.cm} \cinst{1}{Joint Institute for Nuclear Research,
Dubna, Russia} \cinst{2}{Yerevan Physics Institute,
Armenia}\cinst{3}{Institute for High Energy Physics, Protvino,
Russia}
\end{center}
\vspace{100mm}

{\centerline{\bf YEREVAN  2010}}


\newpage
\vspace{1.cm}
\begin{abstract}

For the first time, the nuclear effects in the reaction $\nu p
\rightarrow \mu^- p \pi^+$ on bound protons are investigated at
$\langle E_\nu \rangle \approx$ 9 GeV and the effective atomic
weight $A_\mathrm{eff} \approx 21$ of the composite nuclear target
using the data obtained with SKAT bubble chamber. The observed
nuclear effects are explained by the Fermi-motion of the target
proton and secondary elastic scattering of neutrinoproduced
hadrons on intranuclear neutrons. The probability of any
intranuclear interaction of neutrinoproduced hadrons is extracted:
$P_{int} = 0.40\pm0.13$, while the probability of interactions
occurring only via elastic scattering on intranuclear neutrons is
estimated to be $P_c = 0.14\pm0.05$.

A clear evidence of the $\Delta^{++}(1232)$ and an indication on
the $\Delta^{++}(1620)$ isobar states production are observed with
the mean yields, respectively, $0.57\pm0.09$ and $0.40\pm0.26$
(the latter being corrected for the $p \pi^+$ decay fraction). The
mean yield of $\Delta^{++}(1232)$ is by $1.61\pm0.25$ times
smaller than expected for the reaction on the hydrogen target,
thus indicating on its rather strong nuclear absorption.

\end{abstract}

\newpage
\setcounter{page}{1}
\begin{center}
{\large 1. ~INTRODUCTION}\\
\end{center}


In the space-time pattern of the leptoproduction on nuclear
targets an important role belongs to hadron resonances which cause
a significant fraction of final hadrons (see \cite{ref1} and
references therein). A fraction of short-living resonances (like
$\rho, \Delta$ etc.) decays inside the nucleus followed by
secondary intranuclear interactions of the decay products,
resulting in a deterioration of their effective mass distribution
(i.e. the resonance spectral function). Moreover, the latter can
be distorted also at the propagation of resonances inside the
nuclear medium due to their interactions with intranuclear
nucleons (for example, via the channel $\Delta N \rightarrow N N$)
reducing their lifetime, i.e. widening their spectral function.
The Pauli blocking for the decay of barionic resonances, on the
contrary, leads to the increasing of the resonances lifetime, i.e.
to the decreasing of their width. The nuclear effects can also
induce a  mass shift of resonances (see \cite{ref2,ref3,ref4} and
references therein).

Lepton-induced exclusive reactions on intranuclear nucleons are
convenient processes to observe the nuclear effects in the
resonance production. The aim of this work is the investigation of
nuclear effects in neutrino-induced single-pion production
reaction on intranuclear protons

\begin{equation}
\nu + p \rightarrow \mu^- + p + \pi^+ ,
\end{equation}

\noindent which includes both the $\Delta^{++}(1232)$ production,

\begin{equation}
\nu + p \rightarrow \mu^- + \Delta^{++}(1232) ,
\end{equation}

\noindent and the nonresonant $\pi^+$ production.

To this end, the data from the SKAT bubble chamber \cite{ref5}
were used. In Section 2 the experimental procedure is described.
Section 3 presents the observed kinematic characteristics of the
reaction (1) and their description in terms of nuclear effects.
Section 4 is devoted to the study of the reaction (2), with a
particular emphasis on the nuclear effects influence on the yield
of $\Delta^{++}(1232)$. The results are summarized in Section 4.

\begin{center}
{\large 2. ~EXPERIMENTAL PROCEDURE}\\
\end{center}

The experiment was performed with SKAT bubble chamber, exposed to
a wideband neutrino beam obtained with a 70 GeV primary protons
from the Serpukhov accelerator. The chamber was filled with a
propane-freon mixture containing 87 vol\% propane ($C_3H_8$) and
13 vol\% freon ($CF_3Br$) with the percentage of nuclei H:C:F:Br =
67.9:26.8:4.0:1.3 \% and the effective atomic weight
$A_\mathrm{eff} \approx 21$ of the composite $(C, F, Br)$ nuclear
target. A 20 kG uniform magnetic field was provided within the
operating chamber volume.

Charged (CC) current interactions containing a negative muon with
momentum $p_{\mu} >$0.5 GeV/c were selected. Protons with momentum
below 0.6 GeV$/c$ and a fraction of protons  with momentum
0.6-0.85 GeV$/c$ were identified by their stopping in the chamber.
Stopping $\pi^+$ mesons were identified by their
$\pi^+$-$\mu^+$-$e^+$ decay. A fraction of low-momentum
($p_{\pi^+} < 0.5$ GeV$/c$) $\pi^+$ mesons were identified by the
mass-dependent fit provided that the $\chi^2$- value for the pion
hypothesis was significantly smaller as compared to that for
proton. It was required the errors in measuring the momenta be
less than 24\% for muon, 60\% for other charged particles and
$V^0$'s (corresponding to strange particles) and less than 100\%
for photons. Each event was given a weight to correct for the
fraction of events excluded due to improperly reconstruction. More
details concerning the experimental procedure, in particular, the
reconstruction of the neutrino energy $E_\nu$, can be found in our
previous publications \cite{ref6,ref7}.

Events with $3< E_{\nu} <$ 30 GeV were selected, provided that the
summary momentum of produced hadrons is directed forward
respective to the neutrino direction, resulting in 8237 events (or
$N_\mathrm{tot}$ = 9631 weighted events) with the mean neutrino
energy $\langle E_{\nu} \rangle$ = 9.1 GeV. The contamination from
the neutral current (NC) interactions was estimated to be about
4\%.

We chosen the events-candidates to the reaction (1) containing two
positively charged hadrons and no registered $\gamma$-quanta or
$V^0$- particles. In the most fraction (69\%) of the chosen
events, at least one of the hadrons was identified as $\pi^+$-
meson or proton. The remaining 31\% of events (with two
non-identified hadrons) were weighted by an additional factor of
0.5, because they entered twice in the distributions discussed in
the next sections. The proton (identified or being a unidentified
candidate to proton) is required to be emitted in the forward
direction in the laboratory frame and to have a momentum $P_p
>$0.2 GeV$/c$ in order to reduce the contamination from the
nuclear disintegration products. The problems related with the
background events with misidentified positively charged hadrons
are discussed below in this and next sections.

Further, we rejected events for which the energy momentum
disbalances $p_L^{miss}$ and $p_T^{miss}$ (defined analogously to
those in the next section) were compatible with the kinematics of
the quasielastic reaction $\nu n \rightarrow \mu^- p$ followed by
a secondary elastic $pp$ scattering with at least one
non-identified proton in the final state. The rejected events was
satisfying the requirements $\mid p_L^{miss}\mid < 0.1$ GeV$/c$
and $p_T^{miss} < 0.25$ GeV$/c$, the quoted boundaries being
chosen so that the number of rejected events (19 events) was equal
to that estimated from the differential cross section of the
reaction $\nu n \rightarrow \mu^- p$ and the probability of a
secondary $pp$ scattering resulting in at least one non-identified
proton in the final state.

We also excluded the events containing a registered neutron and
being compatible with the kinematics of the reaction $\nu p
\rightarrow \mu^- \pi^+\pi^+ n$, namely, when the angle
$\varphi_n$ between the neutron direction and the direction of the
summary momentum of the $\mu^-\pi^+\pi^+$ system in the transverse
plane (perpendicular to the neutrino direction) exceeded
143$^\circ$ (i.e. $\cos \varphi_n < - 0.8)$. We kept other events
containing registered neutrons a part of which can originate from
secondary elastic scattering of neutrinoproduced hadrons on
intranuclear neutrons $-$ a process which is one of the subjects
of the present study.

The total number of accepted events-candidates to the reaction
(1), composing a subsample $B_1$, was equal to 802. The
contamination from the NC events is found to be equal to
$2.2\pm0.5$\%.  The contamination from the background CC events
which could contain non-registered neutral particles was estimated
to be $9\pm1$\%, using events with registered neutral particles
weighted by a probability that a given event could potentially
pass the aforementioned selection criteria.

It should be also stressed that a fraction of the subsample $B_1$
corresponds to the exclusive reaction (1) on the hydrogen. The
number $N_p^{free}$ of these events, estimated from the total
number of events $N_{tot}$, the total cross section of $\nu p$ and
$\nu n$ CC interactions \cite{ref8,ref9} and the cross section of
the reaction (1) on free protons \cite{ref8,ref10,ref11} at
$\langle E_\nu \rangle$ = 9.1 GeV, is equal to $N_p^{free} =
115\pm21$. Similarly, the expected event number of the reaction
(1) occurring on bound protons (both followed or not by any
intranuclear secondary interaction) is estimated to be
$N_p^{bound} = 411\pm 65$. The sum of $N_p^{free} + N_p^{bound}$
turns out to be smaller than the number of events in the subsample
$B_1$, contrary to what was expected due to the loss of events
caused by secondary inelastic interactions of neutrinoproduced
proton, $\pi^+$ meson or $\Delta^{++}$. We conclude, therefore,
that the subsample $B_1$ contains a significant fraction of
background events corresponding to a number competing processes
(considered in details in the next section).

\begin{center}
{\large 3. ~THE CHARACTERISTICS OF THE REACTION (1) ON INTRANUCLEAR PROTONS}\\
\end{center}
As compared to the reaction on the hydrogen target, the kinematic
characteristics of the reaction (1) on intranuclear protons suffer
distortions reflecting the effects of the Fermi-motion of the
bound proton and the secondary intranuclear scattering of
neutrinoproduced hadrons. These result in an apparent violation of
the energy-momentum balance of the reaction. Another source of
distortions are experimental uncertainties inherent for both
hydrogen and nuclear events. The energy-momentum unbalance can be
characterized by 'missing' longitudinal ($p_L^{miss}$) and
transverse ($p_T^{miss}$) momenta defined as

\begin{equation}
p_L^{miss} = \sum_{i}{}{(E^i - p_L^i) - m}
\end{equation}

\begin{equation}
{(p_T^{miss})}^2 = (\sum_{i}{}{\overrightarrow{p}_T^i)}^2  \, ,
\end{equation}
where the sums are over energies $E^i$, longitudinal $p^i_L$ and
transverse $\overrightarrow{p}_T^i$ momenta (with respect to the
neutrino direction) of final particles; $m$ is the proton mass.

Figure 1 shows the distributions on $p_L^{miss}$ and $p_T^{miss}$
for the $B_1$ subsample, from which the corresponding
distributions for the background events (see Section 2 above) were
subtracted. As it will be shown below, the characteristic peaks at
low values of $\mid p_L^{miss} \mid$ and $p_T^{miss}$ reflect
mainly the interactions on the hydrogen and, partly, on bound
protons, when final hadrons escape secondary intranuclear
interactions.

In general, the following process can contribute to the plotted
distributions: {\it{a}}) the reaction (1) on hydrogen; {\it{b}})
the reaction (1) on bound protons not followed by any secondary
interaction; {\it{c}}) the reaction (1) on bound protons followed
by elastic scattering of final hadrons or the intermediate
$\Delta^{++}$ on intranuclear neutrons not resulting in the change
of the reaction topology; {\it{d}}) and {\it{e}}) background
reactons followed, respectively, by a charge-exchange secondary
interaction or by a pion absorption on a nucleon pair resulting in
the change of the reaction topology, so that the latter imitates
that of the reaction (1).

The shapes of the $p_L^{miss}$- and $p_T^{miss}$- distributions
corresponding to the processes $a)$-$e)$ are determined by
Monte-Carlo simulations and presented in Figure 1. The
normalization of the depicted curves will be discussed below. The
simulated processes are: \\
{\it {a) The reaction on hydrogen}} \\
The simulation code includes several sources of the experimental
uncertainties: the error in the three-momentum measurement for
final particles, the minor effects related to the neutrino beam
divergence and the muon inner bremsstrahlung (see \cite {ref12}
and references therein), as well as to the particle
misidentification in a fraction of events with two non-identified
positive hadrons (see Section 2 above). The simulated
distributions $f_a(x)$ ($x$ being $p_L^{miss}$ or $p_T^{miss}$),
shown in Figure 1 by dot-dashed curves, describe the sharp peak
near $p_L^{miss} \sim 0$, as well as an essential fraction of the
$p_T^{miss}$- distribution at very low values of $p_T^{miss} < $
0.05 GeV$/c$ corresponding to the reaction (1) on the hydrogen. \\
{\it {b) The Fermi-motion of the bound proton}} \\
In addition to the effects described in item {\it{a)}} above, the
Fermi-motion of the target proton is introduced (according to
\cite{ref13}) leading to a further widening of the distributions
on $p_L^{miss}$ and $p_T^{miss}$. The simulated distributions
$f_b(x)$, as it is seen from Figure 1 (the dashed curves), are
responsible for a prominent fraction of events with
low $\mid p_L^{miss} \mid$ and $p_T^{miss}$. \\
{\it {c) The intranuclear scattering of neutrinoproduced hadrons}}
\\
In addition to the effects described in items {\it{a)}} and
{\it{b)}} above, we included in the simulation code the elastic
scattering of neutrinoproduced hadrons on intranuclear neutrons,
provided that the proton and $\pi^+$ escaped any other
interactions which could lead to the changing of the reaction
topology. The elastic differential $p n$ and $\pi^+ n$ cross
sections and inelastic proton-nucleon and pion-nucleon cross
sections are taken from compilations
\cite{ref14,ref15,ref16,ref17}. A characteristic feature of the
simulated distributions $f_c(x)$, the shapes of which are depicted
by dotted curves, is  their spreading over a wide range of
$p_L^{miss}$ and $p_T^{miss}$ and, for the case of $p_L^{miss}$, a
shifting toward positive values (also observed experimentally).
The reason for this shifting is that in an elastic scattering the
hadron energy $E^i$ decreases on an average in a lesser degree
than its longitudinal momentum $p_L^i$ (cf. Eq. 3). The said
effects are also inherent for the elastic process $\Delta^{++} n
\rightarrow \Delta^{++} n$. The shapes of the distributions
$f_{c'}(x)$ for the latter, inferred from simulations (using the
relevant cross sections predicted theoretically in [3,18,19]),
differ only slightly from $f_{c}(x)$. This difference will be
taken into account when describing the experimental distributions
plotted in Figure 1. \\
{\it {d) The background reactions caused by charge-exchange
intranuclear interactions}}  \\
In addition to the effects described in items {\it{a)}} and
{\it{b)}}, we included in the simulation code several background
processes which could imitate the topology of the reaction (1),
namely: the reaction $\nu n \rightarrow \mu^- n \pi^+$ followed by
the neutron elastic scattering, resulting in a recoil proton in
the final state, and the reaction $\nu p \rightarrow \mu^- p
\pi^0$ followed by the $\pi^0$ charge-exchange reaction $\pi^0 p
\rightarrow \pi^+ n$. The shapes of the simulated distributions
$f_{d'}(x)$ and $f_{d"}(x)$ for these reactions turned out to be
very close to $f_c(x)$, hence, only the summary contribution from
{\it{c)}} and {\it{d)}} can be considered when describing
distributions plotted in Figure 1. We also estimated the possible
contribution from the reaction $\nu n \rightarrow \mu^- p$
followed by an inelastic reaction $p p \rightarrow p n \pi^+$ and
found it to be negligible (about 1\%) in the subsample of
events-candidates to the reaction (1). \\
{\it {e) The background reactions caused by a pion absorption on a
nucleon pair}} \\
We included in the simulation code the reactions $\nu p
\rightarrow \mu^- p \pi^+ \pi^0$ and $\nu p \rightarrow \mu^- n
\pi^+ \pi^+ \pi^0$ followed by $\pi^0$ absorption on a neutron
pair, $\pi^0(n n) \rightarrow n n$, as well as the reactions $\nu
n \rightarrow \mu^- p \pi^+ \pi^-$ and $\nu n \rightarrow \mu^- n
\pi^+ \pi^+ \pi^-$ followed by $\pi^-$ absorption on a
quasideuteron, $\pi^-(p n) \rightarrow n n$ (the details
concerning the pion absorption probabilities can be found in
[24,25] and references therein). Contrary to the processes
{\it{c)}} and {\it{d)}}, the nuclear absorption of a
neutrinoproduced pion leads to a shifting of the $p_L^{miss}$-
distribution towards negative values, because of the omission in
(3) of a positive term ($E^\pi - p_L^\pi$) corresponding to the
absorbed pion. The simulated shapes $f_e(x)$ of the processes
{\it{e)}} are depicted in Figure 1 by thin curves. \\
We fitted together the $p_L^{miss}$- and $p_T^{miss}$-
distributions with four$-$parameter expressions

\begin{equation}
F(x) = N_a \cdot f_a(x) + N_b \cdot f_b(x) + N_{cd} \cdot
f_{cd}(x) + N_e \cdot f_e(x),
\end{equation}

\noindent where $x$ denotes $p_L^{miss}$ or $p_T^{miss}$. All
functions in (5) are normalized to unity. The function $f_{cd}(x)$
is a linear contribution of simulated distributions $f_c(x)$,
$f_{c^\prime}(x)$, $f_{d^\prime}(x)$, $f_{d^{\prime\prime}}(x)$
with relative weights which were varied in wide ranges, in view of
the lack of information on the expected contributions of
subprocesses described in items {\it{c}}) and {\it{d}}). The fit
parameters $N_a$, $N_b$ and $N_e$ correspond to the event numbers
of processes {\it{a}}), {\it{b}}) and {\it{e}}), while $N_{cd} =
N_c + N_d$ is the summary number of events corresponding to the
processes {\it{c}}) and {\it{d}}). The fitted values of these
parameters are equal to $N_a = 100\pm12\pm3$, $N_b =
246\pm36\pm9$, $N_{cd} = 148\pm25\pm13$ and $N_e = 66\pm39\pm21$,
where the first error is statistical and the second one reflects
the uncertainty in the relative weights of the subprocessrs
described in items {\it{c}}) and {\it{d}}). An example of the fit
results is presented in Figure 1. Although the fit quality is not
bad ($\chi^2/ndf \approx 1.8$), the fitted $p_T^{miss}$-
distribution somewhat underestimates the data at relatively high
values of $p_T^{miss} = 0.4-0.75$ GeV$/c$. This discrepancy can
be, at least partly, caused by the fact that our simulation code
uncorporates no more than one intranuclear scattering of a given
hadron.

The inferred value of $N_a$ is in good agreement with the expected
value of $N_p^{free} = 115\pm21$ (cf. Section 2), thus indicating
the self-consistency of our results. The value of $N_b$,
corresponding to the number of events on bound protons (possessing
Fermi-motion) not followed by any secondary interaction, composes
$60\pm13$\% of the expected total number $N_p^{bound} = 411\pm65$
of events occurred initially via the channel (1) on bound protons
(cf. Section 2). Hence, the probability of secondary intranuclear
interactions, both violating or not the topology of the reaction
(1) is equal to $P_{int} = 0.40\pm0.13$. Further, we estimated
from simulations that the probability $P_c$ of the process
{\it{c}}) with respect to the probability $P_{int}$ composes
$P_c/P_{int} = 0.36$, resulting in $P_c = 0.14\pm0.05$ and hence
the event numbers $N_c = P_c \cdot N_p^{bound} = 59\pm27$ and $N_d
= 89\pm38$.

The number $N_c$ along with $N_b$ gives the number of survived
events of the reaction (1) on bound protons, $N_b + N_c =
305\pm33$ composing $74\pm8$\% of $N_p^{bound}$, while the
remaining $26\pm8$\% of events turn out to be rejected due to
secondary interactions violating the reaction topology.

It should be also noted, that the total number of survived events
of the exclusive reaction (1) both occurred on free and bound
protons is equal to $N_{exc} = N_a + N_b + N_c = 405\pm36$, with a
rather small fraction of $N_c = 59\pm27$ events in which the
neutrinoproduced proton and/or $\pi^+$ meson suffered an elastic
scattering on intranuclear neutrons. We conclude, therefore, that
in the case of the $\Delta^{++}(1232)$ production in the exclusive
channel (2) the deterioration of its spectral function due to
these elastic interactions is expected to be faint.

\begin{center}
{\large 4. ~THE $\Delta^{++}(1232)$ EXCLUSIVE PRODUCTION}\\
\end{center}

For the analysis of the $\pi^+ p$ effective mass distribution, the
events with $\mid p_L^{miss} \mid <$ 0.4 GeV$/c$ and $\mid
p_T^{miss} \mid <$ 1 GeV$/c$ were selected. This allow us, as it
follows from Figure 1, to save practically all events of the
reaction (1) and to reject a fraction of events belonging to the
background reactions described in the items $d)$ and $e)$ of the
previous section. The $\pi^+ p$ effective mass distribution is
plotted in Figure 2. Apart from a clear $\Delta^{++}$ signal, a
comparatively narrow peak is seen around 1.6 GeV$/c^2$ which can
correspond to the narrowest excited state of $\Delta^{++}$,
namely, the $S_{31}$ - wave $\Delta^{++}(1620)$ with the
Breit-Wigner mass $\sim 1630$ MeV$/c^2$ and the full width $\Gamma
\approx 145$ MeV (see \cite{ref22}). Other $\Delta^{++}$ states
with masses up to 2 GeV$/c^2$ are too wide ($\Gamma \sim
200\div500 MeV$) and hence cannot be disentangled from the
non-resonance background. Moreover, their signals are expected to
be practically invisible due to their small $\pi^+ p$ decay
fraction \cite{ref22}.

To describe the mass distribution we used a fit function

\begin{eqnarray}
N(m)& = &N(1232) \cdot [(1 -P_c) \cdot BW^{exp}_{1232}(m) +
P_c \cdot B W^{scat}_{1232}(m)] +  \nonumber\\
&& N(1620)\cdot BW^{exp}_{1620}(m)+BG(m) \, \, ,
\end{eqnarray}

\noindent being the sum of two simulated Breit-Wigner functions
\cite{ref23} for $\Delta(1232)$ and $\Delta(1620)$, smeared
according to the experimental inaccuracies, and a smooth
background function assuming that it involves the contributions
both from the non-resonant background inherent to the reaction (1)
and from the background processes described in the items $d)$ and
$e)$ above, as well as the summary contribution from all
large-width ($\Gamma >$ 150 MeV) $\Delta^{++}$ states. The
background function was parametrized by a simplest form $BG(m) =
\alpha m^\beta \exp(-\gamma m)$. Besides, we included in the fit
function the simulated distribution $BW^{scat}_{1232}(m)$
corresponding to the elastic scattering of the $\Delta^{++}(1232)$
decay products on intranuclear neutrons. The probability of the
latter $P_c = 0.14\pm0.05$ was estimated in the previous section.
In this case, a factor $1 - P_c$ was introduced for the
$\Delta(1232)$ spectral function $BW^{exp}_{1232}(m)$ which did
not incorporate scattering effects. Due to the smallness of the
$\Delta(1620)$ contribution to the $\pi^+ p$ mass distribution and
of the scattering probability $P_c$, we neglected the scattering
effects in its spectral function $BW^{exp}_{1232}(m)$. Note, that
all Breit-Wigner functions in (6) are normalized to unity.

An example of the fit result at $P_c = 0.14$ is plotted in Figure
2. The thin solid and dotted curves correspond to the
$\Delta(1232)$ production followed or not by secondary scattering
processes with probabilities, respectively, $P_c = 0.14$ and $1 -
P_c$ = 0.86. The dot-dashed and dashed curves correspond to the
$\Delta(1620)$ production and the background distribution. The sum
of these four contributions is depicted by the thick solid curve
which satisfactorily describes the experimental data (with
$\chi^2/ndf = 1.2)$. It should be noted, that the limited
statistics of our data does not allow to infer any conclusion
concerning the mass shift and the widening of the
$\Delta^{++}(1232)$ resonance caused by its propagation in the
nuclear medium \cite{ref2,ref3,ref4}.

The fit results in the numbers $N(1232)$ of produced
$\Delta^{++}(1232)$, (both followed or not by secondary elastic
scattering of the decay products) and $N(1620)$ of
$\Delta^{++}(1620)$ (for the decay mode of $\pi^+ p$) being equal
to $N(1232) = 230\pm30\pm8$ and $N(1620) = 40\pm25\pm1$, where the
quoted errors reflect, respectively, the statistical uncertainty
and the uncertainty in $P_c = 0.14\pm0.05$. The corresponding mean
yields, normalized to the event number $N_{exc} = 405\pm36$ of the
exclusive reaction (1), are equal to $\bar{n}(1232) = 0.57\pm0.09$
and $\bar{n}^{corr}(1620) = 0.40\pm0.25$, the latter being
corrected for the $\pi^+ p$ decay fraction $(25\pm5\%)$
\cite{ref22}.

It should be noted that the mean yield of $\Delta^{++}(1232)$
underestimates the cross section ratio $r_{exp} = \sigma(\nu p
\rightarrow \mu^- \Delta^{++})/\sigma(\nu p \rightarrow \mu^- p
\pi^+)$ deduced from available experimental data on free protons,
$r_{exp} = 0.85\pm0.14$ at $\langle E_\nu \rangle$ = 3 GeV
\cite{ref24} and $r_{exp} = 0.79\pm0.16$ at $\langle E_\nu
\rangle$ = 27 GeV \cite{ref25}, as well as from theoretical
calculations at $\langle E_\nu \rangle \sim$ 9 GeV: $r_{theor}
\approx$ 0.92 (see \cite{ref10,ref25} and references therein). We
conclude, therefore, that the nuclear absorption effects are more
prominent in the channel (2) as compared to the channel (1). This
'extra' absorption can be quantified by a factor
$r_{theor}/\bar{n}(1232)= 1.61\pm0.25$. The observed suppression
of the $\Delta^{++}(1232)$ yield can be, at least partly,
explained by its large absorption cross section via the channel
$\Delta^{++} n \rightarrow p p$ on intranuclear neutrons
\cite{ref2,ref3,ref26}.

\begin{center}
{\large 5. ~SUMMARY}\\
\end{center}

For the first time, the nuclear effects influencing the kinematic
characteristics of the reaction $\nu p \rightarrow \mu^- p \pi^+$
(1) on bound protons are studied and explained by the Fermi-motion
of the target proton and secondary elastic scattering of
neutrinoproduced hadrons on intranuclear neutrons. The probability
of any intranuclear interaction of neutrinoproduced hadrons is
extracted: $P_{int} = 0.40\pm0.13$, while the probability of
interactions occurring only via elastic scattering on intranuclear
neutrons is estimated to be $P_c = 0.14\pm0.05$.

A clear evidence of the $\Delta^{++}(1232)$ and an indication on
the $\Delta^{++}(1620)$ isobar states production in the reaction
(1) are observed, with the mean yields $\bar{n}(1232) =
0.57\pm0.09$ and $\bar{n}(1620) = 0.40\pm0.26$ (the latter being
corrected for the $\Delta^{++}(1620)\rightarrow p \pi^+$ decay
fraction). The mean yield of $\Delta^{++}(1232)$ is by
$1.61\pm0.25$ times smaller than expected for the reaction (1) on
the free proton, thus indicating on its noticeably nuclear
absorption before its decay.

\begin{center}
{\large ACKNOWLEDGMENTS}\\
\end{center}

The authors from YerPhI acknowledge the supporting grants of
Calouste Gulbenkian Foundation and Swiss Fonds Kidagan. The
activity of one of the authors (H.G.) is supported by Cooperation
Agreement between DESY and YerPhI signed on December 6, 2002.


\newpage
\begin{figure}[ht]
\resizebox{0.9\textwidth}{!}{\includegraphics*[bb =20 65 600
610]{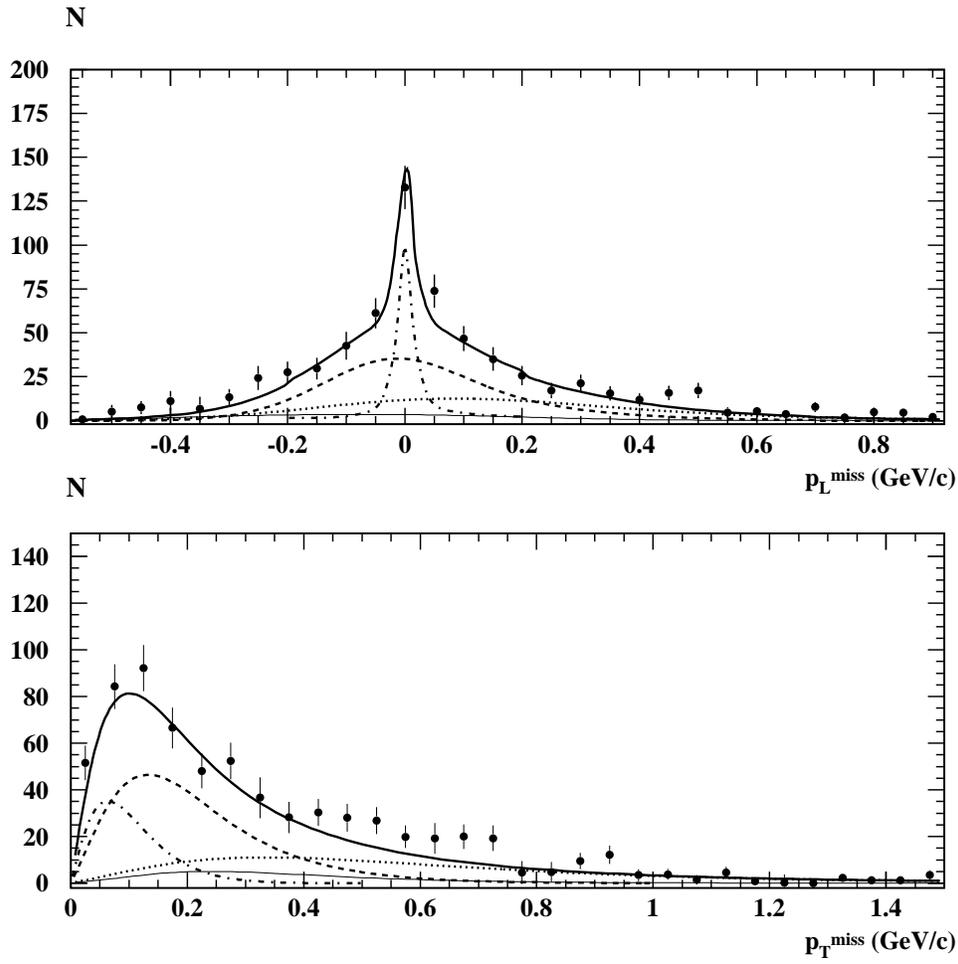}} \caption{The $p_L^{miss}$ - and $p_T^{miss}$ -
distributions. The curves are the fit results (see the text).}
\end{figure}

\newpage
\begin{figure}[ht]
\resizebox{0.9 \textwidth}{!}{\includegraphics*[bb=20 40 500 610]
{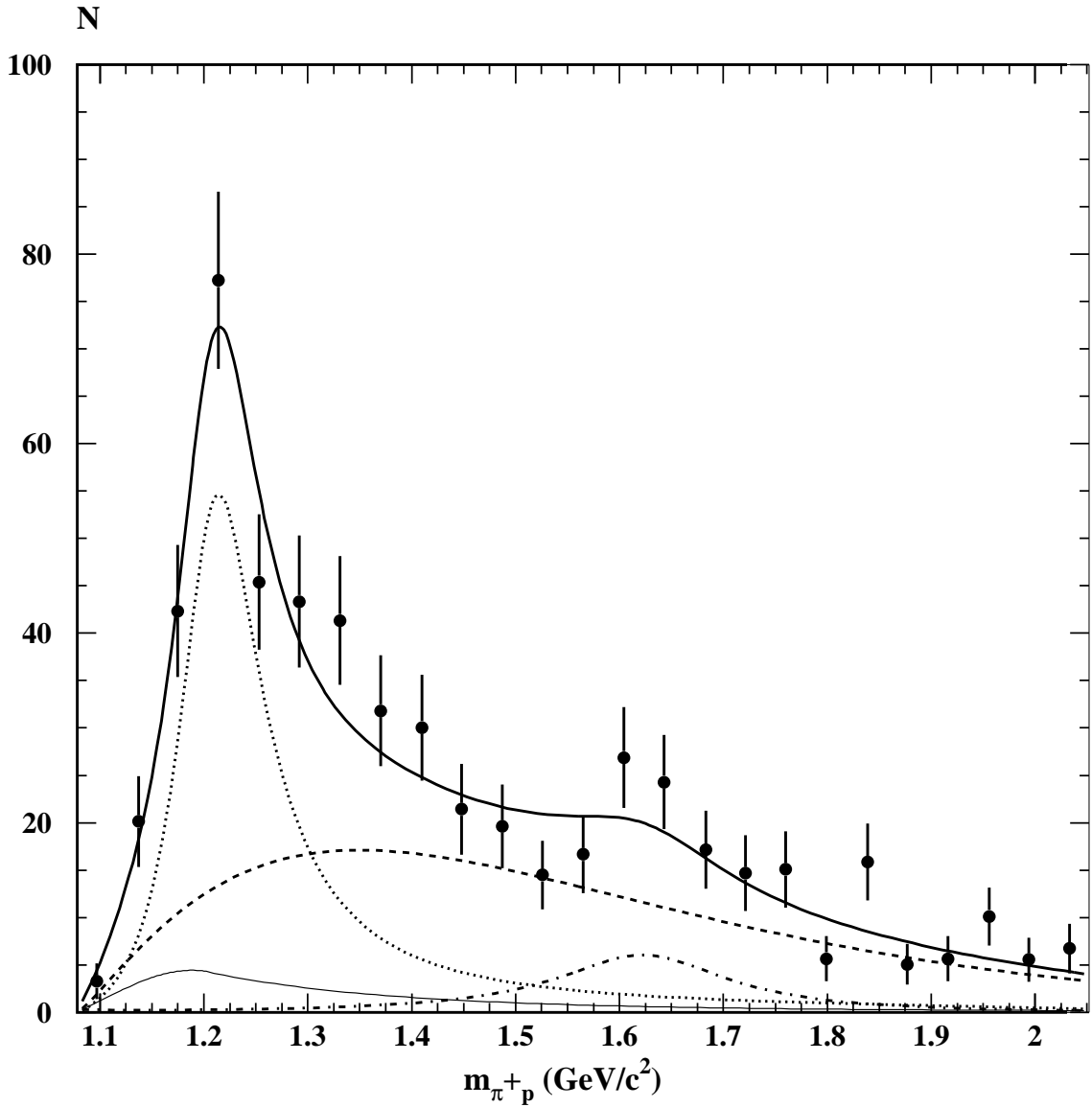}} \caption{The $\pi^+ p$ effective mass distribution.
The curves are the fit results (see the text).}
\end{figure}

\end{document}